\documentclass[longbibliography,superscriptaddress,twocolumn,showpacs,preprintnumbers,amsmath,amssymb,nofootinbib,nobibnotes,aps,prstper]{revtex4-1}
\usepackage{graphicx}% Include figure files
\usepackage{dcolumn}% Align table columns on decimal point
\usepackage{bm}% bold math
\usepackage{rotating}% Allow table entries to be rotated
\usepackage{multirow}

\begin{document}

\title{The Design and Validation of the {Q}uantum {M}echanics {C}onceptual {S}urvey}


\author{S. B. McKagan}
\altaffiliation[Now at: ]{American Association of Physics Teachers and McKagan Enterprises, Seattle, WA, 98144, USA}
\affiliation{JILA, University of Colorado and NIST, Boulder, CO 80309, USA}

\author{K. K. Perkins}
\affiliation{Department of Physics, University of Colorado, Boulder, CO 80309, USA}

\author{C. E. Wieman}
\affiliation{Department of Physics, University of British Columbia, Vancouver, BC V6T 1Z1, CANADA}
\affiliation{JILA, University of Colorado and NIST, Boulder, CO 80309, USA}
\affiliation{Department of Physics, University of Colorado, Boulder, CO 80309, USA}

\date{\today}

\begin{abstract}
The Quantum Mechanics Conceptual Survey (QMCS) is a 12-question survey of students' conceptual understanding of quantum mechanics.  It is intended to be used to measure the relative effectiveness of different instructional methods in modern physics courses.  In this paper we describe the design and validation of the survey, a process that included observations of students, a review of previous literature and textbooks and syllabi, faculty and student interviews, and statistical analysis.  We also discuss issues in the development of specific questions, which may be useful both for instructors who wish to use the QMCS in their classes and for researchers who wish to conduct further research of student understanding of quantum mechanics.  The QMCS has been most thoroughly tested in, and is most appropriate for assessment of (as a posttest only), sophomore-level modern physics courses.  We also describe testing with students in junior quantum courses and graduate quantum courses, from which we conclude that the QMCS may be appropriate for assessing junior quantum courses, but is not appropriate for assessing graduate courses.  One surprising result of our faculty interviews is a lack of faculty consensus on what topics should be taught in modern physics, which has made designing a test that is valued by a majority of physics faculty more difficult than expected.
\end{abstract}

\maketitle

\section{Goals of the Survey}
The Quantum Mechanics Conceptual Survey (QMCS) was designed to be used as a general survey of students' conceptual understanding of quantum mechanics.  Our primary goal was to create a survey that could be used to measure the relative effectiveness of different curricula and teaching techniques, and thus to serve as a formative assessment tool for faculty engaging in reform of their courses.  To achieve this goal it is important to focus on standard topics that are covered in most courses, both traditional and reformed.  Therefore, the QMCS does not cover many topics that we believe are important but are not universally taught, including measurement, applications of quantum mechanics, and historical experiments.  Further, to have an impact on faculty practice, the test questions must be written such that the average faculty member believes that they are measuring something important.  A secondary goal in creating the survey was to provide a way to measure the prevalence of specific student difficulties.  As we will discuss, we have found that this secondary goal is often in conflict with the primary goal.

The QMCS was originally intended to be applicable to all levels of quantum mechanics, from sophomore-level modern physics to graduate quantum mechanics.  This intention was based on the assumption that there is a set of universally agreed-upon basic concepts that each level of quantum mechanics instruction builds upon at a deeper level.  However, we have found that this is not a particularly good assumption, and we were not able to design a test that is truly applicable at all levels.  The QMCS is most relevant to, and has been most thoroughly tested in, sophomore-level modern physics courses.  In these courses, we recommend giving the QMCS as a posttest at the end of a course only, and not as a pretest at the beginning of the course.  As we will discuss, most students have little or no knowledge of any of the concepts measured by the QMCS at the beginning of a modern physics course, so it is both meaningless and demoralizing as a pretest.

The QMCS has also been used in junior-level and graduate-level quantum courses.  However, faculty generally regard the test as too basic for these courses.  In graduate courses, there is no difference between pretest and posttest scores, even for students who score low on the pretest, and there is no correlation between QMCS scores and final exam scores.  These results are consistent with faculty assertions that they do not teach the concepts measured by the QMCS in graduate courses.  However,  graduate students in selective programs do score higher on the QMCS than do physics graduate students in less selective programs.

In keeping with standard practice for research-based conceptual tests in physics~\cite{Redish2003a}, we have protected the security of the test by keeping it from becoming available to students.  After administering it in class we do not post solutions or allow students to take the tests home.  We have administered it online to graduate students and advanced undergraduates, but always on a password-protected site with a warning not to distribute it.  To further protect the security of the test, we do not include it in this article.  It can be accessed from our website~\cite{QMCS} by requesting a password from the authors.

\section{Review of Previous Research}
There is a long history of research-based conceptual tests of specific topics in physics, starting with the Force Concept Inventory (FCI)~\cite{Hestenes1992a}.  Widespread use of the FCI has played a particularly important historical role in spreading awareness of the deficiencies in traditional instruction in physics~\cite{Hake1998a}.  While many of the finer details of its validation have been criticized~\cite{Huffman1995a,Heller1995a}, defended~\cite{Halloun1995a,Halloun1995b,Henderson2002a}, and studied~\cite{Steinberg1997a,McCullough2001a,Rebello2004a,Dancy2006a}, this test has several important strengths:
\begin{enumerate}
  \setlength{\itemsep}{1pt}
  \setlength{\parskip}{0pt}
  \setlength{\parsep}{0pt}
    \item It is written in everyday language without jargon or mathematics, so that students do not need to be familiar with a particular notational convention and can easily understand the questions even before taking a physics class.
    \item It focuses on conceptual understanding rather than standard problems so students cannot rely on memorized algorithms to answer the questions.
    \item Many of the distracters are very effective at eliciting students' preconceived ideas about the world and thus reveal the extent to which student learning in class has been superficial.
    \item Most faculty initially believe that the questions are ``too easy,'' so they are startled when their students do badly (as often happens) and therefore motivated to examine their teaching.
\end{enumerate}
The FCI has been used to measure the relative effectiveness of curricula in many physics courses throughout the world and has had a major impact on educational practices in introductory physics.

Many other research-based tests have been developed for introductory mechanics and other topics in physics.  Some of these tests, like the FCI, are ``concept inventories'' that attempt to measure student understanding of a single narrowly defined concept~\cite{Hestenes1992a,Beichner1994a,Thornton1998a,Engelhardt2004a}.  Other tests are ``surveys'' that attempt to measure student understanding of a broad range of topics within a subject~\cite{Maloney2001a,Singh2003a,Ding2006a}.  The QMCS falls into the latter category, but its range is still only a fraction of what is covered in a modern physics course.
%concept inventories: FCI,TUG-K, FMCE, DIRECT
%surveys: CSEM, EMCS, BEMA

Several research-based tests in quantum mechanics have already been developed, including the Quantum Mechanics Visualization Instrument (QMVI)~\cite{Cataloglu2002a,Cataloglu2002b}, the Quantum Mechanics Concept Inventory (QMCI)~\cite{Falk2004a}, and the Quantum Physics Conceptual Survey (QPCS)~\cite{Wuttiprom2006a}.  However, all these tests serve different functions than the QMCS.  While the authors of the QMVI claim that it is a general test of quantum mechanics, most of the questions focus on a single concept: the relationship between the shape of the wave function and the shape of the potential.  While this is an important concept, it is not all there is to quantum mechanics, and it is possible to understand this concept quite well without having any conceptual understanding of how quantum mechanics describes reality.  Further, the QMVI is extremely difficult, with many questions on topics that are not typically covered until graduate school, if at all, and we have known PhDs in physics to score less than 80\% on it.  The QMCI is designed to be a concept inventory rather than a survey, focusing on the concepts of ``one-dimensional potential barriers, tunneling and probability distribution,'' according to its author.  For these particular concepts, it is much more thorough than the QMCS.  The QPCS also has a narrower focus, on the concepts of wave-particle duality and the photoelectric effect.

In addition to these multiple-choice tests, Singh has developed surveys with open-ended questions in junior-level~\cite{Singh2001a} and graduate-level~\cite{Singh2008a} quantum mechanics.  However, the purpose of these surveys appears to be mainly for research in determining student difficulties in quantum mechanics, rather than for comparing instructional techniques among different courses.

While some guidelines for developing conceptual tests in physics have been published~\cite{Lindell2007a,Adams2010a}, there is no consensus in the physics education research community about how conceptual tests should be validated.  Many different validation techniques have been used, criticized, and defended.  Some authors focus on validation through student interviews~\cite{Thornton1998a,Falk2004a}, others focus on statistical techniques~\cite{Maloney2001a,Cataloglu2002a,Ding2006a}, and others use some combination~\cite{Singh2003a,Engelhardt2004a,Lindell2007a,Adams2010a}.

One issue that has been largely ignored in validation of conceptual tests is faculty buy-in.  This turns out to be very important for a conceptual test in quantum mechanics.  For a test to impact faculty practice in the broader physics community, rather than just being an internal tool of the physics education research community, average faculty members must believe that it is testing something they value for their students.  Thus, tests that focus on identifying obscure misconceptions that faculty members don't recognize may be convincing to researchers in PER, but will not have a significant impact on faculty practice.  This is widely recognized in the development of surveys of student beliefs, where test designers interview physics faculty members in order to validate the ``expert'' view~\cite{Redish1998a,Adams2006a}, but tends to be ignored in the development of conceptual tests, where test designers are more likely to assume that they know the expert view.  As discussed below, we found there is large variation in faculty views on many topics in quantum mechanics.

In addition to research on the development of conceptual tests, we have drawn on research on student understanding of quantum mechanics.  Previous research has explored many student difficulties in understanding quantum mechanics, including difficulty with understanding the relationship of amplitude and wavelength to potential~\cite{Ambrose1999a,Bao1999a,Sadaghiani2005a} and with drawing correct wave functions for tunneling plane waves~\cite{Ambrose1999a,Bao1999a,Morgan2004a,Wittmann2005a}.  Research has also uncovered many specific incorrect beliefs common among students, such as that particles move along sinusoidal paths~\cite{Steinberg1996a,Muller2002a,Olsen2002a,Knight2004b,Falk2004a}, that energy is lost in tunneling~\cite{Ambrose1999a,Bao1999a,Morgan2004a,Wittmann2005a,Falk2004a,Domert2005a,McKagan2006a} , and that reflection and transmission are due to particles having a range of energies~\cite{Ambrose1999a,Bao1999a,Domert2005a}.  QMCS questions designed to address these issues are discussed in Section VB.

\section{Design of the QMCS}
A preliminary version of the QMCS was developed with questions based on initial ideas gathered from several sources: faculty interviews, a review of textbooks and syllabi, observations of students in lectures and study sessions, and a literature review of known student difficulties.

\subsection{Faculty Interviews}
Before writing any questions for the QMCS, we interviewed 8 faculty members (7 physics and 1 chemistry), most of whom had recently taught a course in modern physics, and asked them to list ``the three most important concepts you would like your students to get out of a course in modern physics.''  The goal of these interviews was to develop a list of learning goals based on faculty consensus about what should be taught in modern physics, and then to develop QMCS questions to address each of these goals.  However, the most important result of our faculty interviews is that there is \emph{not} a faculty consensus about what should be taught in modern physics.  Some faculty members responded that their primary goal in this class was not to teach concepts at all, but to teach calculational skills, history, or applications.  Some faculty members listed concepts, but these concepts often did not overlap.

We grouped the concepts that did show some overlap into common themes, listed below, with the number of faculty members who gave a response related to each theme in parentheses:
\begin{itemize}
  \setlength{\itemsep}{1pt}
  \setlength{\parskip}{0pt}
  \setlength{\parsep}{0pt}
\item wave function and probability (5)
\item wave-particle duality (4)
\item Schrodinger Equation (4)
\item quantization of states (4)
\item uncertainty principle (3)
\item superposition (3)
\item operators and observables (3)
\item tunneling (2)
\item measurement (2)
\end{itemize}
We note that only the first item on the list was given by more than half the faculty interviewed.  While this list does not reflect the most important goals of many faculty, it is the closest we could come to a consensus on the most important topics to be covered in a modern physics.  This list was used as a general guide in developing the QMCS, and we attempted to write questions to address each item in the list.

The last item in the list, measurement, was controversial.  Some faculty listed measurement as one of the most important concepts in modern physics, and the authors consider it to be critical to understanding the connection between quantum mechanics and reality.  However, several other faculty specifically said that measurement is \emph{not} important and should not be taught in modern physics courses.  Because we wanted the QMCS to focus on topics that are universally taught, this result, along with the results discussed below from our review of textbooks and syllabi, led us to exclude this topic from later versions of the QMCS.  While some questions do indirectly address measurement, there are no questions in Version 2.0 with measurement as a primary learning goal.

\subsection{Review of textbooks and syllabi}
We reviewed textbooks and syllabi in modern physics and quantum mechanics to see what topics in modern physics are most important to experts.  We found that textbooks had a great deal of overlap in terms of the types of examples and calculations they included, a result that makes sense given the limited number of solvable problems in quantum mechanics.  There was much less overlap in terms of the concepts included and the degree to which conceptual understanding was emphasized at all.  

Further, most textbooks included little or no discussion of the issues most frequently raised in the physics education research literature on quantum mechanics.  For example, most textbooks do not discuss whether particles move in sinusoidal paths, whether energy is lost in tunneling, or how to determine the qualitative shape of a wave function from a potential.

One fact we found particularly surprising is that not a single modern physics textbook we reviewed included any discussion of measurement, and only about half of introductory quantum mechanics discussed measurement at all.  This is consistent with our results from faculty interviews that many faculty do not consider measurement to be an important concept.

Our results on measurement are consistent with results reported by Dubson et al.~\cite{Dubson2009a}.  In their review of junior quantum textbooks, they found that 30\% did not mention the wave function collapse and 40\% made ``only brief, passing mention'' of it.  They also found, in faculty interviews, that 11\% of faculty thought that measurement should not be taught in junior quantum (they did not ask about modern physics), and an additional 26\% ``were deeply troubled by the idea of wave function collapse... and consequently teach it with some unease.''

\subsection{Observations of students}
Before implementing any course reforms or developing a survey, the first author observed lectures and ran an informal study session for students in an introductory modern physics course for engineering majors.  The optional study session was advertised to students as an opportunity to work together and get help on their homework.  The first author helped students with questions about any topic in the course, including homework and studying for exams, interacting with them directly and observing them talking to each other.  She took notes about student questions and comments in lectures and study sessions, and shared these notes with the other authors.  These notes were an important source of information about student ideas and difficulties with the course material, and informed both the development of the QMCS and the reforms that we implemented in the course the following semester.  We wrote questions to address specific difficulties observed in lectures and study sessions, and based the wording of answers, both correct and incorrect, on quotes from students.

Study sessions and observations were continued in reformed courses in subsequent years, both by the authors and by undergraduate learning assistants who wrote weekly field notes about their interactions with students.  Continuing observations were used to inform the refinement of the QMCS.  We note that observations in lecture were much more informative in reformed classes in which students engaged in frequent small group discussions in response to clicker questions than in more traditional lecture courses because there were more opportunities to hear students thinking out loud.

\subsection{Addressing known student difficulties}
In addition to writing questions to address the topics that faculty believe are more important, we wrote questions to address known student difficulties, both from our own observations discussed in the previous section, and from the literature discussed in Section II.  Many questions on the QMCS are either adapted, or taken directly, from questions in existing literature on student difficulties. For example, question 10 is adapted from the QMVI, and was included based on extensive research showing that students have difficulty relating the variance in local wavelength of a wave function to the shape of the potential~\cite{Ambrose1999a,Bao1999a,Ambrose2005a}.  Question 11 is taken directly from the QMVI.  Question 7 and an earlier version of it are based on extensive research showing that students often believe that energy is lost in tunneling.~\cite{Ambrose1999a,Bao1999a,Morgan2004a,Wittmann2005a,Falk2004a,Domert2005a,McKagan2006a}  Question 5 is also taken from a survey used in previous research.~\cite{Fletcher2004a}

\subsection{Design challenges specific to Quantum Mechanics}
There are several issues that make it significantly more difficult to design a conceptual test in advanced topics such as quantum mechanics than in introductory mechanics.  Thus, while we attempted to design a survey that has the strengths of the FCI, some of these strengths are not attainable.

One strength of the FCI is that it does not contain jargon and is comprehensible even to a student with no physics background.  While it is possible to include more or less jargon in a test of quantum mechanics, one cannot avoid all reference to wave functions, energy levels, probability densities, and potential energy diagrams, concepts which have little meaning to students unfamiliar with quantum mechanics.  This is true not only in quantum mechanics, but in many other areas of physics, such as electromagnetism, where students' low scores on pretests are often due more to a lack of knowledge of the subject than to incorrect beliefs.

Students also have far fewer preconceived ideas about quantum mechanics than about the topics covered in introductory mechanics.  While students certainly have preconceived ideas about the world that quantum mechanics contradict (e.g. that a particle can have a well-defined position and momentum at the same time), they do not have preconceived ideas about wave functions and many of the entirely new concepts that are introduced in quantum mechanics.  While one might imagine that students have learned about quantum mechanics through popular culture and movies such as \emph{A Brief History of Time} and \emph{What the \#\$*! Do We Know!?}, we were surprised by how infrequently students referred to any prior knowledge of quantum mechanics in interviews on the QMCS.  In fact, one student, a 35-year-old engineering major, told us he had never heard that electrons could behave as waves before taking our course.  The physics majors we interviewed usually had more awareness of quantum mechanics in popular culture than the engineering majors.

One result of students' lack of prior knowledge of quantum mechanics is that, at least in a modern physics course, giving the QMCS as pretest is both meaningless and demoralizing.  We began giving the QMCS as both a pretest and a posttest and calculating normalized gains mainly because this is standard practice for research-based conceptual tests.  However, no matter how much we emphasized to students that it was OK if they didn't know the answers and they would not be graded, they often told us after taking the pretest that it was discouraging and scary.  Due to this observation, and the low pretest scores discussed in the next section, we do not recommend giving the QMCS as a pretest in modern physics courses.

A more serious problem in developing a conceptual survey of quantum mechanics is that the very notion of a conceptual understanding of the this subject is controversial.  By ''conceptual understanding," we mean an understanding of the relationship between the mathematical description and the physical world, an understanding of what is actually happening, rather than just the ability to calculate a result.  We emphasize conceptual understanding because we believe it is the basis of a true understanding of any subject.  However, we recognize that this view of quantum mechanics is not universally accepted, with prominent physicists urging their students to "shut up and calculate," and even arguing that the best one can hope for in quantum mechanics is to be able to do a calculation and predict an experimental result and that understanding what is actually happening is not possible.

Much of what is typically taught in this subject involves algorithmic problem-solving skills that do not have an underlying conceptual base.  For example, many faculty have the goal that their students should be able to solve the Schrodinger equation for specific abstract potentials such as a harmonic oscillator or an infinite square well.  However, it is easy to memorize the method or solution for such specific problems, and difficult to apply this method to an entirely new problem.  One standard technique for testing conceptual understanding in introductory physics is to put concepts in new contexts, so that students cannot rely on memorized patterns from contexts they have already studied.  However, because there are so few solvable problems in quantum mechanics, it is difficult to find contexts that are likely to be new for all students.  Further, most problems involve sufficiently complex mathematics that a conceptual understanding is not sufficient to solve most new problems.

A great deal of quantum mechanics involves specific ``tricks'' and techniques that are easy if one has seen them before, but nearly impossible to think of on one's own.  For example, question 16 on the QMVI requires using symmetry to solve for the energies of a ``half-well,'' a technique that is simple enough if one has learned it, but is not always taught.  We were careful to avoid questions that tested whether students had learned a particular technique or method.

A further problem, discussed above, is that there is a lack of faculty agreement on what should be taught, especially at the modern physics level.  Even when faculty used the same words to describe their goals in interviews, what they meant by those words was not always clear, and they did not necessarily value questions designed to address their stated goals.  We also found that faculty did not value questions designed to elicit known student difficulties, such as the belief that particles move along sinusoidal paths, because they did not believe that such conceptual issues were relevant to solving problems in quantum mechanics.  Thus, our two goals of comparing instruction that would be convincing to faculty and determining the prevalence of known difficulties were sometimes at odds.

Because of the problems discussed in this section, the final version of the QMCS contains only 12 questions, significantly fewer than we believe is ideal.  Previous versions of the QMCS have contained up to 25 questions, but many of the original questions have been eliminated due to problems uncovered in student interviews, graduate student surveys, and discussions with faculty.  Examples of problems with eliminated questions are discussed in Section V.  Even the remaining 12 questions still have some problems (also discussed in Section V), but they are the questions that we have found to provide the most useful insight into student thinking.

The current version of the QMCS has too few questions to adequately probe student understanding of quantum mechanics.  Further, the small number of questions leads to a discretization of possible scores and other problems with statistical analysis.  Developing new questions is a project for future research, beyond the scope of this paper.  Louis Deslauriers and Carl Wieman have developed and validated four additional questions for the QMCS, which will be described in a future publication.

\section{Refinement and Validation of the QMCS}

After developing an initial version of the QMCS as discussed in the previous section, we refined and validated it using student interviews, expert review, and finally, statistical analysis.  This was an iterative process extending over approximately three years, in which we developed many different versions\footnote{Old versions are available at http:/per.colorado.edu/QMCS.}, which were administered in interviews to dozens of students and in class to hundreds of students in many different courses.

\subsection{Student Interviews}
We have conducted interviews on various versions of the QMCS with 49 students.  With the exception of one graduate student and four engineering majors who had completed modern physics the previous semester, all interviews were conducted either during the last week of a course in modern physics or quantum mechanics (see Table \ref{courses} for a breakdown of the courses), or shortly after the students completed the course.  In most cases, students had already completed the survey in class before the interview (in the remainder, they saw it for the first time in the interview).  In the interviews, students were asked to talk through each question, explaining out loud which answer they had picked and why.  The interviewer prompted the students to speak out loud if they were quiet and occasionally asked clarifying questions.  More in-depth questions about students' thinking were saved for the end of the interview so that they would not affect students' answers to later survey questions.  Each interview lasted about an hour, although a few enthusiastic students stayed up to two hours.  Students were paid \$20 for their time.  This allowed us to recruit students with a wider range of abilities, rather than only the ``good'' students who would be willing to volunteer their time for an interview for free.

\begin{table}
\caption{\label{courses}Courses from which students were drawn for interviews.  N is the number of students interviewed from each course.  The numbers for the first three courses include students from several different semesters of the same course.}
\begin{tabular}{|l|r|}
\hline
Course & N \\
\hline
Modern physics for engineering majors (Reformed) & 24 \\
Modern physics for engineering majors (Traditional) & 8 \\
Modern physics for physics majors (Traditional) & 15 \\
Junior quantum mechanics (Traditional) & 1 \\
Graduate quantum mechanics (Traditional) & 1 \\
\hline
Total & 49\\
\hline
\end{tabular}
\end{table}

These interviews helped us to refine the wording of questions, understand student thinking, and eliminate some questions that were not serving their intended purpose.  Some specific results of interviews have been discussed elsewhere.~\cite{McKagan2006a,McKagan2008c}  Further results about student thinking about quantum mechanics will be discussed in Section V.  Examples of general issues about survey design uncovered in interviews include:
\begin{itemize}
  \setlength{\itemsep}{1pt}
  \setlength{\parskip}{0pt}
  \setlength{\parsep}{0pt}
    \item Students tend to avoid any response that includes words like ``exactly,'' ``definitely,'' or ``certainly,'' because ``nothing is ever definite in quantum mechanics.''  We have eliminated such words from the final version of the QMCS to ensure that students' responses are based on their understanding of the underlying concepts rather than on the wording.
    \item Students often skim and miss the word ``not'' in negatively worded questions or answers.  We have changed such questions to use positive wording where possible, and italicized the word ``not'' where it is unavoidable.
    \item In early versions, many questions contained the option ``I have no idea how to answer this question,'' to remove the noise of students who were guessing.  However, most students were unwilling to select this option even when they frankly admitted in interviews that they had no idea how to answer the question, explaining that they preferred to guess than to admit that they didn't know.  When the survey was given as a posttest in class, fewer than 10\% of students chose this option for any question.  The few students who did pick this option in interviews often did so because they were not completely sure of their answer, even when they were able to make a reasonable educated guess.  We have eliminated this option from all questions because whether students select it appears to be a function of personality rather than of whether they actually understand the question.
\end{itemize}

\subsection{Faculty validation}

After writing questions, we asked faculty members teaching courses in which we wished to give the survey to review it and give us feedback about the survey and how they expected their students to do on it.  One important result from these discussions is that we did not fully succeed in our goal of making a survey like the FCI in that all faculty believe that it covers concepts that their students should (and do) know.  The questions were designed to address concepts that many faculty listed as important, and most students are able to interpret the questions on the current version correctly.  However, many faculty still do not necessarily value the questions, either because they do not think the concepts addressed by the questions are important, or because they misinterpret them even though students do not.  Further, most faculty recognize that there are enough subtle issues in quantum mechanics that they do not expect their students to do particularly well on the QMCS, and thus are not shocked by the results in the way that they are by the results of the FCI.

We also attempted to get faculty feedback by sending out an online version of the survey asking for general feedback and ranking of the importance of questions to all physics faculty members at a large research university and to a listserv for faculty members in PER.  This yielded only 3 responses, all of which were fairly atypical compared to the faculty we interviewed.  We suspect that in addition to being very busy, faculty are unwilling to respond to such a survey for fear of getting the answer wrong.  In fact, we have seen that a surprising number of faculty get some questions wrong because they do not have a correct understanding of the relevant concept.  On the other hand, physics graduate students had no such qualms and were thus able to serve as an effective substitute for faculty for the ``expert'' response.  We sent the online survey to all graduate students at the same university and listserv and received 27 responses from physics grad students, including detailed comments, within a month of sending the email.  Almost half the responses came within 3 hours.  Many of these responses were from grad students at other universities that did not have PER groups, indicating that grad students forwarded the survey to their friends.

Grad students wrote extensively in the optional comment boxes on the online survey, which were very helpful in refining the wording of questions to make them more palatable to experts.  The grad students often answered questions ``incorrectly'' for very different reasons than modern physics students.  For example, in one question on measurement, which has since been eliminated because the topic is not valued by most faculty, nearly half the grad students answered incorrectly because they were worrying about decoherence, a topic we had not intended to test.  Questions which were too hard for most grad students to answer correctly were either modified or eliminated.  For example, as will be discussed in Section VB2, only 15\% of grad students answered the original version of question 10 (see Fig.~\ref{slanted}) correctly, and the question was modified to make it easier.

We have found that it is impossible to write questions to which experts cannot find obscure objections, and thus have focused on writing questions that are unambiguous to students and for which most experts can agree on the answer and importance.  Even in question 1, one of the least controversial questions on the QMCS, one expert has pointed out that because the probability of emission is proportional to the energy spacing, increasing the spacing would lead to more photons being an emitted from an ensemble, so E could be a correct answer.  Such objections are difficult to address without adding a long list of caveats that tend to confuse students.  Because we have never heard this reasoning from students, or even from more than one expert, we have left the question as it is.

\subsection{Statistical Analysis}

\begin{figure}[htbp]
  \includegraphics[width=\columnwidth]{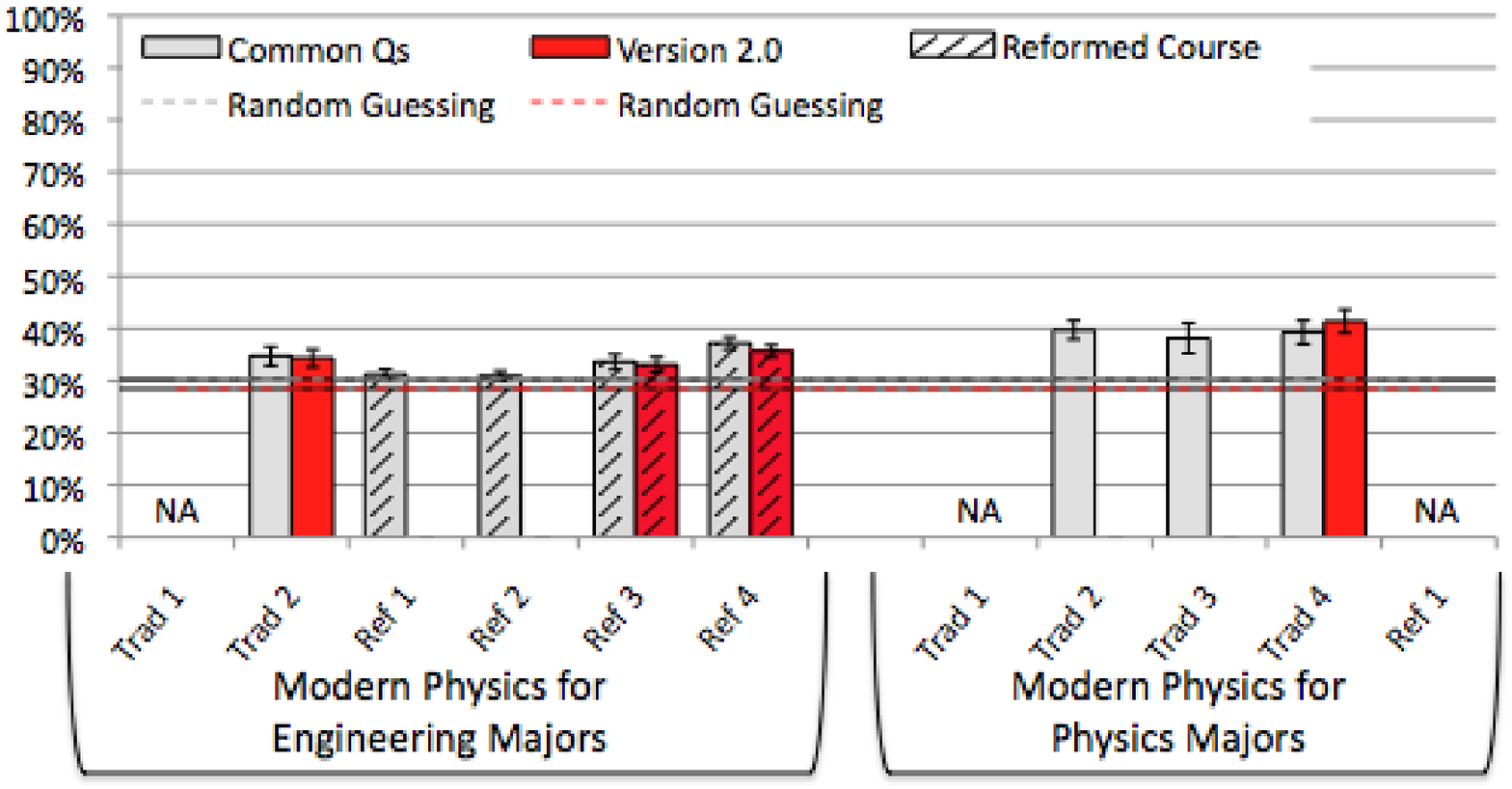}
  \caption{\label{pretest}Average pretest scores in different modern physics courses for common questions (gray) and Version 2.0 questions (red).  Solid colored bars indicate courses taught with traditional instruction and cross-hatched bars indicate reformed courses.  Error bars show standard error on the mean.  The two dashed horizontal lines show the average score that would be expected if students answered using random guessing for the common questions (gray) and Version 2.0 (red).  The chart includes some courses in which no pretest was given in order to compare more easily to Fig.~\ref{posttest}.}
\end{figure}

\begin{figure*}[bhpt]
  \includegraphics[width=\textwidth]{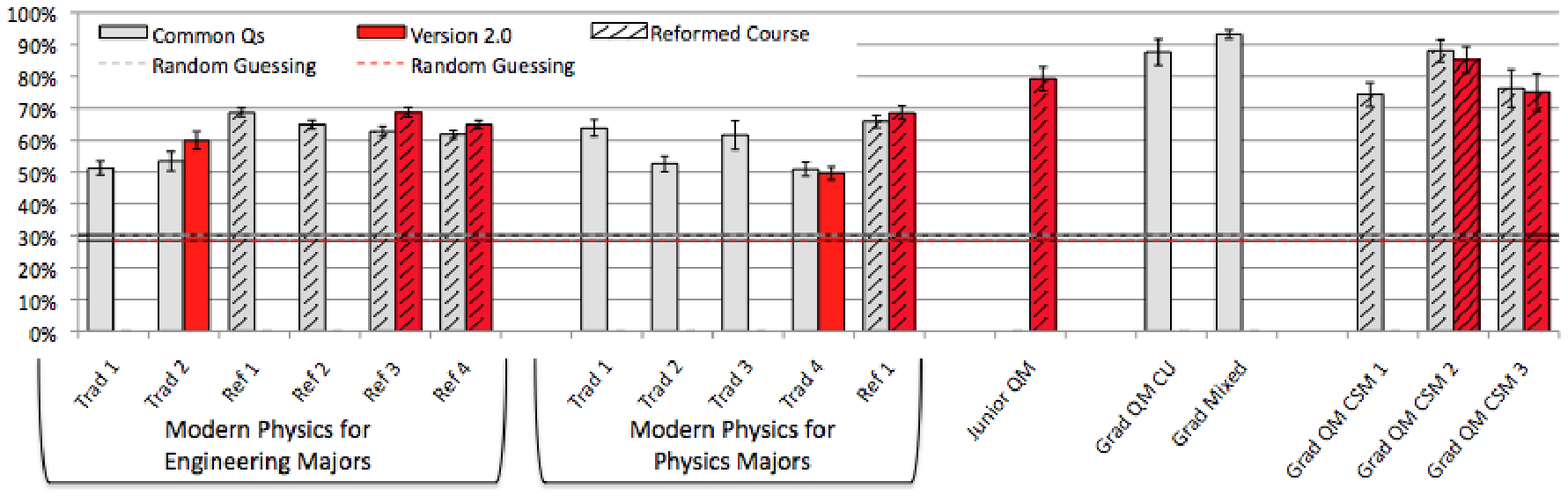}
  \caption{\label{posttest}Average posttest scores in different courses for common questions (gray) and Version 2.0 questions (red).  Solid colored bars indicate courses taught with traditional instruction and cross-hatched bars indicate reformed courses.  Error bars show standard error on the mean.  All scores are post-test scores except for ÒGrad QM CUÓ, which is from a pre-test for an introductory graduate quantum course at CU, and ÒGrad mixedÓ, which is from a survey given over the summer to graduate students from different universities who may or may not have taken a graduate quantum course.}
\end{figure*}

Tables \ref{predata} and \ref{postdata} and Figs. \ref{pretest} and \ref{posttest} show the average pretest and posttest scores, respectively, for two versions of the QMCS in different modern physics courses.  The gray bars show average scores for the 12 common questions that were asked in all earlier versions of the test.   The red bars show average scores for questions on the current version of the test, Version 2.0.  The 12 common questions include nine questions that are included in Version 2.0 and three additional questions that have been eliminated.\footnote{The three eliminated questions include the two questions about reflection and transmission at a potential step shown in Fig.~\ref{RT}, which were eliminated because we found that many faculty do not cover this topic in modern physics, and one question asking whether photons travel in sinusoidal paths, which was eliminated because we did not want to have two questions about a single topic that many faculty found of questionable value.  See Section VB for discussion of eliminated questions.} Three of the questions in Version 2.0 (7, 10, and 12) were added relatively late in the development of the test.  Therefore, we have much more data for the common questions than for Version 2.0 questions.

\begin{table}[htbp]
\caption{\label{predata}Average pretest scores in different modern physics courses for common questions asked all semesters and for Version 2.0 questions.  V is the version number given in each course.  We list the total number of students enrolled at the beginning of the semester, along with the number of students who took the QMCS pretest in each course.  Uncertainties are standard errors on the mean.  Courses in bold were given a version that included all questions in Version 2.0 and are thus included in the analysis of Version 2.0 in Section IVC3.}
\begin{tabular}{|l|l|l|r|r|r|r|}
\hline
\multicolumn{2}{|c|}{\multirow{2}{*}{\textbf{Course}}} & \multirow{2}{*}{\textbf{V}} & \multicolumn{2}{|c|}{\textbf{\# students}} & \multicolumn{2}{|c|}{\textbf{Pretest Scores}} \\ \cline{4-7}
\multicolumn{2}{|c|}{} & \multicolumn{1}{|c|}{}
& \multicolumn{1}{c|}{Total} & \multicolumn{1}{c|}{QMCS} & \multicolumn{1}{c|}{Common} & \multicolumn{1}{c|}{V 2.0} \\ \hline
Modern  & Trad 1 & 1.0 & \multicolumn{ 4}{c|}{No Pretest Given} \\ \cline {2-7}
Physics & \textbf{Trad} 2 & 1.9 & 69 & 60 & $35\%\pm2\%$ & $34\%\pm2\%$ \\ \cline {2-7}
for & Ref 1 & 1.6 & 191 & 205 & $31\%\pm1\%$ & \multicolumn{1}{c|}{NA} \\ \cline {2-7}
Engi- & Ref 2 & 1.7 & 184 & 189 & $31\%\pm1\%$ & \multicolumn{1}{c|}{NA} \\ \cline {2-7}
neering & \textbf{Ref 3} & 1.8 & 94 & 97 & $34\%\pm1\%$ & $33\%\pm1\%$ \\ \cline {2-7}
Majors & \textbf{Ref 4} & 1.9 & 156 & 149 & $37\%\pm1\%$ & $36\%\pm1\%$ \\ \hline
 &  &  &  \multicolumn{4}{c|}{} \\ [-1em]\hline
Modern  & Trad 1 & 1.2 &  \multicolumn{ 4}{c|}{No Pretest Given} \\ \cline {2-7}
Physics & Trad 2 & 1.6 & 80 & 60 & $40\%\pm2\%$ & \multicolumn{1}{c|}{NA} \\ \cline {2-7}
for & Trad 3 & 1.7 & 38 & 46 & $38\%\pm3\%$ & \multicolumn{1}{c|}{NA} \\ \cline {2-7}
Physics & \textbf{Trad 4} & 1.8 & 69 & 63 & $39\%\pm2\%$ & $41\%\pm2\%$ \\ \cline {2-7}
Majors & \textbf{Ref 1} & 1.9 &  \multicolumn{ 4}{c|}{No Pretest Given} \\ \hline
\end{tabular}
\end{table}

\begin{table}[htbp]
\caption{\label{postdata}Average posttest scores in different courses for common questions asked all semesters and for Version 2.0 questions.  V is the version number given in each course.  We list the total number of students enrolled at the beginning of the semester, along with the number of students who took the QMCS posttest in each course.  Uncertainties are standard errors on the mean.  Courses in bold were given a version that included all questions in Version 2.0 and are thus included in the analysis of Version 2.0 in Section IVC3 and IVC4.}

\begin{tabular}{|l|l|l|r|r|r|r|}
\hline
\multicolumn{2}{|c|}{\multirow{2}{*}{\textbf{Course}}} & \multirow{2}{*}{\textbf{V}} & \multicolumn{2}{|c|}{\textbf{\# students}} & \multicolumn{2}{|c|}{\textbf{Posttest Scores}} \\ \cline{4-7}
\multicolumn{2}{|c|}{} & \multicolumn{1}{|c|}{}
& \multicolumn{1}{c|}{Total} & \multicolumn{1}{c|}{QMCS} & \multicolumn{1}{c|}{Common} & \multicolumn{1}{c|}{V 2.0} \\ \hline
Modern & Trad 1 & 1.0 & 80 & 68 & $51\%\pm2\%$ & \multicolumn{1}{c|}{NA} \\ \cline {2-7}
Physics & \textbf{Trad 2} & 1.9 & 67 & 42 & $53\%\pm3\%$ & $60\%\pm3\%$ \\ \cline {2-7}
for & Ref 1 & 1.6 & 189 & 169 & $69\%\pm1\%$ & \multicolumn{1}{c|}{NA} \\ \cline {2-7}
Engi- & Ref 2 & 1.7 & 182 & 160 & $65\%\pm1\%$ & \multicolumn{1}{c|}{NA} \\ \cline {2-7}
neering & \textbf{Ref 3} & 1.8a & 94 & 78 & $63\%\pm2\%$ & $69\%\pm2\%$ \\ \cline {2-7}
Majors & \textbf{Ref 4} & 1.9 & 153 & 124 & $62\%\pm1\%$ & $65\%\pm1\%$ \\ \hline
 & \multicolumn{1}{r|}{} &  &  &  & & \\ [-1em]\hline
Modern &Trad 1 & 1.2 & 77 & 66 & $64\%\pm3\%$ & \multicolumn{1}{c|}{NA} \\ \cline {2-7}
Physics & Trad 2 & 1.6 & 71 & 66 & $53\%\pm2\%$ & \multicolumn{1}{c|}{NA} \\ \cline {2-7}
for & Trad 3 & 1.7 & 43 & 29 & $61\%\pm4\%$ & \multicolumn{1}{c|}{NA} \\ \cline {2-7}
Physics & \textbf{Trad 4} & 1.8a & 65 & 57 & $51\%\pm2\%$ & $50\%\pm2\%$ \\ \cline {2-7}
Majors & \textbf{Ref 1} & 1.9 & 76 & 69 & $66\%\pm2\%$ & $69\%\pm2\%$ \\ \hline
\multicolumn{2}{|r|}{} & \multicolumn{1}{r|}{} &  &  & & \\ [-1em]\hline
\multicolumn{2}{|r|}{Junior QM} & 2.0\footnote{The test given in the Junior QM course was Version 2.0 plus three additional questions.} & 34 & 22 & \multicolumn{1}{c|}{NA} & $79\%\pm4\%$ \\ \hline
\multicolumn{2}{|r|}{} & \multicolumn{1}{r|}{} &  &  & & \\ [-1em]\hline
\multicolumn{2}{|r|}{Grad QM CU\footnote{The test was given as a pretest only in the Grad QM course at CU, but we have included it here rather than in Table II so that the data can be more easily compared to the other data for graduate students.  Previous research has shown that there is no difference between pretest and posttest scores for graduate quantum courses.}} & 1.6 & 30 & 18 & $88\%\pm4\%$ & \multicolumn{1}{c|}{NA} \\ \hline
\multicolumn{2}{|r|}{Grad Mixed} &1.5 & NA & 27 & $93\%\pm1\%$ & \multicolumn{1}{c|}{NA} \\ \hline
\multicolumn{2}{|r|}{} & \multicolumn{1}{r|}{} &  &  & & \\ [-1em]\hline
\multicolumn{2}{|r|}{Grad QM CSM 1} & 1.6 & 22 & 22 & $74\%\pm4\%$ & \multicolumn{1}{c|}{NA} \\ \hline
\multicolumn{2}{|r|}{Grad QM CSM 2} & 1.9 & 9 & 9 & $88\%\pm3\%$ & $85\%\pm4\%$ \\ \hline
\multicolumn{2}{|r|}{Grad QM CSM 3} & 1.9 & 7 & 7 & $76\%\pm6\%$ & $75\%\pm6\%$ \\ \hline
\end{tabular}
\end{table}

We report the average scores for both the common questions and Version 2.0 questions to give a better sense of the variance of scores between semesters than one could get from the average scores for the Version 2.0 questions alone.  Figs. \ref{pretest} and \ref{posttest} show that, for courses in which we have data for both the common questions and Version 2.0 questions, the scores for the two versions are close, so that while one cannot make a rigorous quantitative comparison between the two versions, it is reasonable to compare scores from the two versions to get a rough sense of the kinds of scores that are typical in various courses.

We note that in previous publications where we have reported QMCS results~\cite{McKagan2007a,Carr2009a}, these results have been for the common questions shown here in gray.  In these publications, because we were interested in looking at gains for individual courses, we reported scores for matched pre- and post- tests.  In the current paper, because we are most interested in looking at pre- and post- tests individually, and because we did not give pretests in all courses, we report the average scores of all students who took the pre-test and the post-test respectively.  Because the data are not matched, the pre- and post- test scores cannot be compared directly to each other or to the data reported in previous publications.

\subsubsection{Pretest Analysis}

As shown in Fig.~\ref{pretest}, modern physics students' pretest scores are only slightly higher than one would expect from random guessing.  Students in the physics majors' course score slightly higher on the pretest than students in the engineering majors' course, a result that is consistent with the general belief among physics faculty that the physics majors are stronger students.

The slight variance from random guessing does not appear to be caused by a significant prevalence of prior knowledge or by attractive misconceptions in the distractors.  Students do not score particularly well on any single question on the pretest, including those such as question 1 about which students might be expected to have some prior knowledge from a high school physics or chemistry course.  They score slightly above random guessing on all questions, with the exception of question 9, on which the majority of students incorrectly answer that there are no allowed energy values so the overall score is worse than random guessing on this question.  This question is not one on which students could be expected to have any prior knowledge or ideas, and the error can be explained by a literal interpretation of the graph: there are no allowed values inside the well because the total energy is above the well, rather than inside it.  The question on which students score the highest above random guessing is question 11, on which students cannot be expected to have any prior knowledge, but on which the correct answer can be determined using test-taking skills.

We do not report pretest scores for junior quantum courses because we do not have them.  We do not report pretest scores for graduate quantum courses because, as we have reported elsewhere~\cite{Carr2009a}, they are indistinguishable from posttest scores, even for students with low pretest scores.

\subsubsection{Posttest Analysis}

More advanced students tend to score higher than less advanced students, with typical scores of 50-70\% for students after modern physics courses, ~80\% for students after a junior-level quantum course, and 75-90\% for graduate students.  When the QMCS is given pre and post, as we have reported elsewhere, students make significant gains in modern physics courses~\cite{McKagan2007a}, and zero gains in graduate quantum courses~\cite{Carr2009a}.  More research is needed to determine whether the difference in average scores between modern physics students and graduate students is due to a selection effect or due to learning later in the undergraduate curriculum (presumably in junior quantum).  In the one junior quantum course in which we have given the QMCS, there was an abbreviated pretest with only six questions.  There was a significant gain (0.43) on these six questions, which suggests that students learned concepts tested by the QMCS in this course.  However, this was a limited subset of questions and an atypical course including many uncommon teaching reforms~\cite{Goldhaber2009a}, so more research is needed to determine whether this is a generalizable result.

The University of Colorado offers two modern physics courses, one for engineering majors and one for physics majors.   The general belief among CU faculty is that the physics majors are stronger students.  This belief is supported by slighter higher pretest scores on the QMCS common questions for the physics majors (38-40\%) than for the engineering majors (31-37\%).  However, there is not much difference between the post-test scores for the two classes shown in Fig.~\ref{posttest}.

Fig.~\ref{posttest} shows scores from a variety of courses that were taught using traditional methods (solid-colored bars) and courses using reformed teaching methods (hatch-marked bars).  The reformed teaching methods are described elsewhere for the modern physics courses~\cite{McKagan2007a}, junior quantum course~\cite{Goldhaber2009a}, and graduate quantum courses~\cite{Carr2009a}.  Early results suggested that reformed instruction in modern physics leads to higher scores on the QMCS~\cite{McKagan2007a}.  After collecting more data, it is less clear that this is the case.  As can be seen in Fig.~\ref{posttest}, the scores for all reformed courses are higher than the scores for all traditional courses of the same type, but on average these differences are small.  Students in reformed courses do score much higher than students in traditional classes on some questions, with the largest differences on questions 4, 9, and 12.  Ref.~\onlinecite{Deslauriers2010a} reports much more dramatic differences in overall QMCS scores between reformed and traditional instruction at the University of British Columbia.

\subsubsection{Comparison of items on pre- and post- tests}

\begin{figure}[htbp]
  \includegraphics[width=\columnwidth]{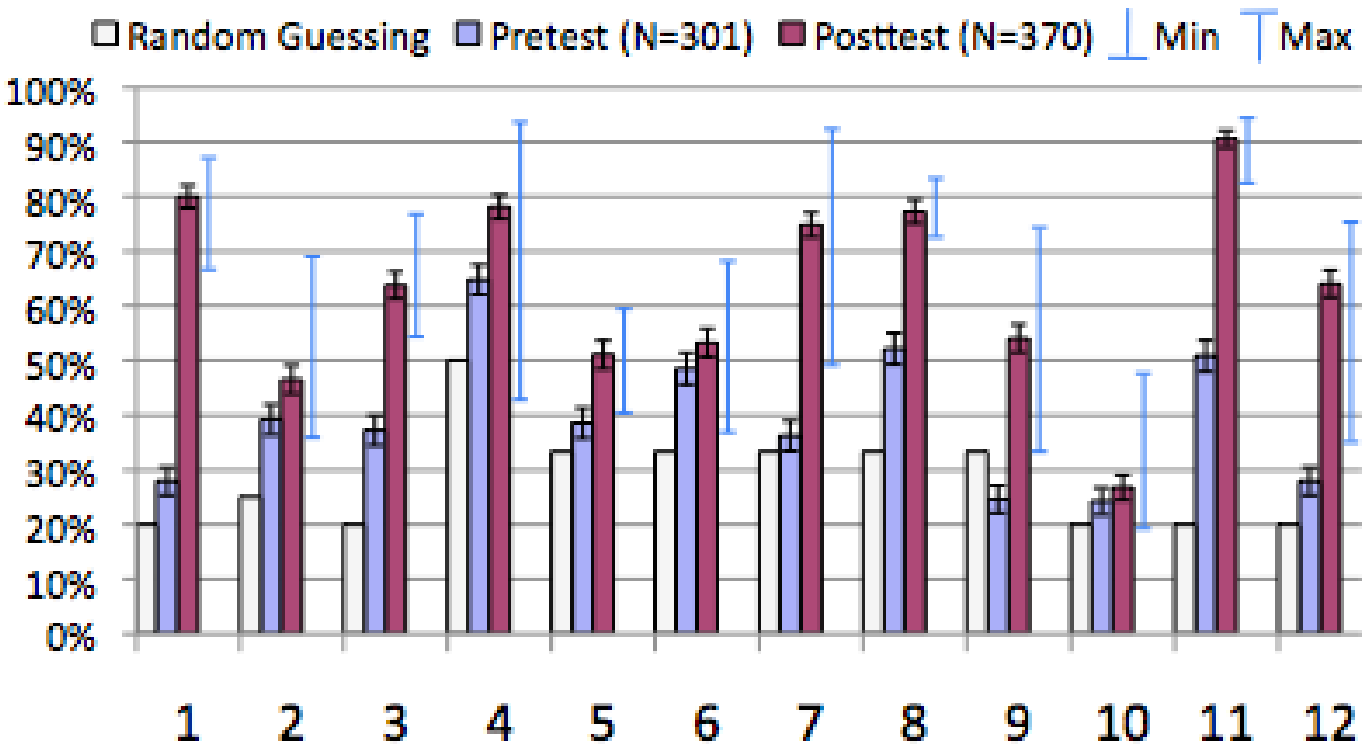}
  \caption{\label{prepost}Scores on individual questions in Version 2.0 in modern physics classes.  This graph combines data from all five modern physics classes in which all Version 2.0 questions were given.  The blue bars on the right show the minimum and maximum average posttest score for individual courses.}
\end{figure}

Fig.~\ref{prepost} shows the scores on individual questions in Version 2.0 in modern physics classes on the pretest and the posttest.  

We note that there is a large variation in gain across questions, with students making large gains on some questions, and little or no gains on others.  In particular, we note that there is zero gain on question 6, about the uncertainty principle, and on question 10, about the qualitative relationship between wavelength and potential.  These results suggest that students learn very little of what is measured by these questions in modern physics courses.

Some questions also have a large variation in posttest scores among courses, as shown by the blue bars.  This range demonstrates that QMCS questions are measuring differences between courses and that different courses are achieving the learning goals measured by those questions to different extents. Examining the different teaching approaches used in the course which achieved the lowest and highest scores on each question would be informative, but is beyond the scope of this paper.

\subsubsection{Item and Test Statistics for Version 2.0}
We have computed item and test statistics that are commonly used as measures of validity and reliability of multiple-choice tests.  For a more detailed discussion of the meaning and significance of these measures and their use in research-based tests in physics and other disciplines, see Refs. ~\onlinecite{Adams2010a} and ~\onlinecite{Ding2006a} and references therein.

For the remainder of the statistical analysis, we have combined posttest data from all modern physics courses in which we have given all questions in Version 2.0, that is, the five courses with red bars in the first two categories in Fig.~\ref{posttest}.  We average over all students, rather than over courses, because we are most interested in overall results.

\begin{table}[htbp]
\caption{\label{responses}Modern physics students' posttest responses to each question in Version 2.0.  Correct answers are in bold.}
\begin{tabular*}{.75\columnwidth}{@{\extracolsep{\fill}}|r|r|r|r|r|r|}
\hline
\multicolumn{1}{|r|}{Question} & \multicolumn{1}{l|}{A} & \multicolumn{1}{l|}{B} & \multicolumn{1}{l|}{C} & \multicolumn{1}{l|}{D} & \multicolumn{1}{l|}{E} \\ \hline
1 & 2\% & 3\% & 6\% & \textbf{80\%} & 9\% \\ \hline
2 & 39\% & 1\% & \textbf{46\%} & 14\% &  \\ \hline
3 & 2\% & 6\% & 18\% & \textbf{64\%} & 9\% \\ \hline
4 & 22\% & \textbf{78\%} &  &  &  \\ \hline
5 & 30\% & \textbf{51\%} & 19\% &  &  \\ \hline
6 & 30\% & \textbf{53\%} & 17\% &  &  \\ \hline
7 & 2\% & \textbf{75\%} & 23\% &  &  \\ \hline
8 & 8\% & \textbf{77\%} & 15\% &  &  \\ \hline
9 & 14\% & 31\% & \textbf{54\%} &  &  \\ \hline
10 & 6\% & 18\% & 34\% & \textbf{27\%} & 14\% \\ \hline
11 & 3\% & 2\% & 2\% & 1\% & \textbf{91\%} \\ \hline
12 & 2\% & 12\% & 14\% & \textbf{64\%} & 5\% \\ \hline
\end{tabular*}
\end{table}

Table~\ref{responses} shows the percentage of modern physics students who give each response for each question on Version 2.0.  This table gives a sense of the difficulty of each question as well as the prevalence of student ideas represented by the distracters.

\begin{table}[htbp]
\caption{\label{stats}Item statistics for Version 2.0 questions given on posttest to modern physics students.  N=370.}
\begin{tabular*}{.55\columnwidth}{@{\extracolsep{\fill}}|c|c|c|c|}
\hline
\begin{sideways}Question\end{sideways} & \multicolumn{1}{l|}{\begin{sideways}Item Difficulty\end{sideways}} & \multicolumn{1}{l|}{\begin{sideways}Discrimination\end{sideways}} & \multicolumn{1}{l|}{\begin{sideways}Pt. Biserial Coeff.\end{sideways}} \\ \hline
1 & 0.80 & 0.45 & 0.47 \\ \hline
2 & 0.46 & 0.46 & 0.39 \\ \hline
3 & 0.64 & 0.56 & 0.47 \\ \hline
4 & 0.78 & 0.47 & 0.47 \\ \hline
5 & 0.51 & 0.44 & 0.38 \\ \hline
6 & 0.53 & 0.50 & 0.36 \\ \hline
7 & 0.75 & 0.41 & 0.37 \\ \hline
8 & 0.77 & 0.19 & 0.19 \\ \hline
9 & 0.54 & 0.46 & 0.36 \\ \hline
10 & 0.27 & 0.28 & 0.26 \\ \hline
11 & 0.91 & 0.19 & 0.29 \\ \hline
12 & 0.64 & 0.54 & 0.47 \\ \hline
& & &\\[-1em]\hline
avg. & 0.63 & 0.41 & 0.37 \\ \hline
\end{tabular*}
\end{table}

Table~\ref{stats} shows selected item statistics for Version 2.0 questions for modern physics students.  The item difficulty, or percent correct, gives a general sense of the difficulty of each item.  (Item difficulty is also shown by the purple bars in Fig.~\ref{prepost} and the bold percentages in Table~\ref{responses}.)  The QMCS contains questions with wide range of difficulties, from 0.27 for the most difficult question to 0.91 for the easiest question, with most questions falling in a range between 0.5 and 0.8.

Item discrimination is a measure of how well an item discriminates between strong students (those who do well on the test as a whole) and weak students (those who do not).  It is calculated by taking the difference between the average scores of the top-scoring students and the bottom-scoring students.  There are various percentages reported in the literature for how many students should be included in each group, ranging from 21\% to 50\%.  We have chosen to use a common method of taking the difference in scores between the top and bottom 27\% of students.  There are various cutoffs reported in the literature for the minimum discrimination an item should have, with the most common being 0.3.~\cite{Doran1980a}  However, Adams and Wieman point out that less discriminating items may serve useful purposes on a test designed to measure course effectiveness.~\cite{Adams2010a}  The majority of items on the QMCS have discrimination $>0.4$, with the exception of three items with particularly low discrimination, questions 8, 10, and 11.  The reasons for the low discrimination of questions 10 and 11 will be discussed in Section V.

The Pearson point biserial coefficient measures the consistency between an item and the test as a whole.  A low point biserial coefficient indicates that student understanding of the concept measured by an item is not correlated to understanding of other concepts on the test.  For tests designed to measure a single concept, low point biserial coefficients are undesirable and the literature says that they should be $\geq0.2$.~\cite{Kline1986a}  However, as Adams and Wieman point out~\cite{Adams2010a}, for a test designed to measure multiple concepts, such as the QMCS, a low point biserial coefficient is not necessarily bad; rather it is an important indicator of a concept that is more difficult to learn than others.  All questions on the QMCS have point biserial coefficients greater than the arbitrary cutoff of 0.2 with the exception of question 8.

Reliability statistics commonly calculated for an entire test include the Kuder-Richardson reliability index (which, if questions are graded as only correct or incorrect, is equivalent to the Cronbach Alpha coefficient), and Ferguson's delta.

The Kuder-Richardson reliability index, or Cronbach Alpha, is a measure of the overall correlation between items.~\cite{Kuder1937a,Cronbach1951a}  It is commonly stated that should be above the cutoff value of 0.7 for an instrument to be reliable.~\cite{Doran1980a}  However, Adams and Wieman argue that instruments designed to measure multiple concepts may have low Cronbach Alpha because these concepts may be independent.~\cite{Adams2010a}  This is the case for the QMCS, which has a Cronbach Alpha of 0.44.

Ferguson's delta is an indication of the discrimination of the test as a whole, as measured by the distribution of scores in the sample.~\cite{Ferguson1949a}  It is commonly stated that for a test to have good discrimination, Ferguson's delta should be $\geq0.9$.~\cite{Kline1986a}  Ferguson's delta for the QMCS is 0.93.

\section{Validity issues with specific questions}
In discussions with faculty and graduate students who are interested in using the QMCS, both for teaching and for research, we have found that many of the issues that cause problems for students are not obvious to experts.  In many cases, the questions that are most appealing to experts may not be the best questions for testing student understanding, either because they are not testing conceptual understanding, or because they are testing multiple concepts, or because students may interpret them in a way that experts do not expect.  In this section, we discuss some of the issues that have come up in developing specific questions, including some questions that have been eliminated from Version 2.0.  We hope that this discussion will aid instructors in interpreting their students' results and in modifying the QMCS appropriately for different instructional environments, and will aid researchers in using QMCS questions for research and in designing new questions to test students' conceptual understanding of quantum mechanics.

\subsection{Questions designed to address important concepts}
The QMCS contains many questions that were designed to address concepts that experts believe to be important.  It has been difficult to design such questions for many reasons, including a lack of conceptual basis for the ideas that experts most value and a lack of faculty agreement on interpretation.

\subsubsection{Relationship of probability density to wave function}
\begin{figure}[htbp]
  \includegraphics[width=\columnwidth]{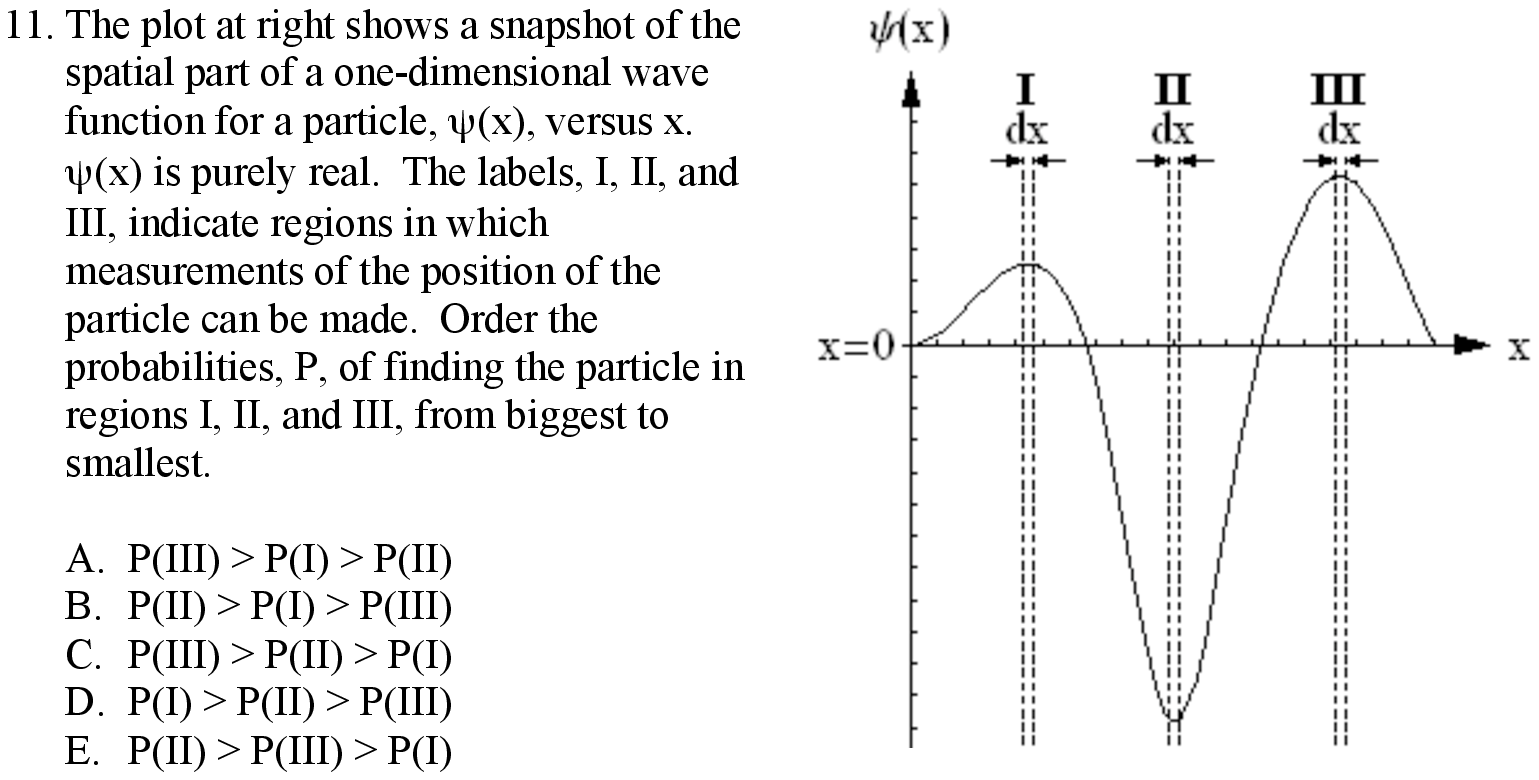}
  \caption{\label{Q11}Question 11, taken from the QMVI, tests student understanding of the relationship between wave function and probability density.}
\end{figure}

The majority of faculty we interviewed stated that understanding wave functions and probability is one of the most important concepts for students in modern physics.  Question 11, shown in Fig.~\ref{Q11}, tests student understanding of the relationship between wave function and probability density.  Most faculty who have reviewed the survey have commented that this is a good question, and it is the question most like a typical homework or exam question.

However, this question is very easy, with 91\% of students answering it correctly, and it shows little discrimination between strong and weak students.  One can answer it simply by memorizing the statement, ``the probability is the square of the wave function,'' without any understanding of what is meant by probability or the wave function.  In interviews with students currently enrolled in modern physics courses, nearly all recited something close to this phrase, circled the correct answer, and quickly moved on to the next question.  The most time-consuming part of answering the question was simply searching through the responses to find the one that matched the answer they had already determined.

Out of four interviews conducted with students who had completed a modern physics course five months earlier, two struggled to remember whether the probability density was related directly to the wave function, or to the absolute value or square of the wave function.  We never observed any such confusion among the remaining interview students who were either currently enrolled in modern physics or had completed it within the last two weeks.  On the other questions on the survey, the four students who had completed the course much earlier had no more trouble than the rest of the students.  While the sample size is small, these results suggest that students may memorize this relationship because it is repeated so often during the course, but do not understand its significance well enough to retain it in long-term memory.

Question 11 is also the only question on which students score more than $20\%$ higher than random guessing on the pretest; $51\%$ of students answer this question correctly on the pretest, $31\%$ more than would be expected if students were guessing randomly.  Since it is highly unlikely that students would have learned the relevant concepts before starting a modern physics course, these results suggest that many students are able to use other reasoning skills to arrive at a correct answer without any understanding of the relevant physics.  Since there are only two quantities mentioned in the problem, the most plausible guess is that a higher magnitude of one gives a higher magnitude of the other.  The only other plausible guess is that it is the values, rather than the magnitudes, that are related, which would lead to answer A.  This was the second most common answer on the pretest, selected by $21\%$ of students.  It is also possible to distinguish between these two plausible guesses without knowing any quantum mechanics simply by knowing that probability cannot be negative.

In spite of these issues, we have retained question 11 because it tests a concept that faculty consider very important and because students make significant gains on it between the pretest and posttest, which shows that they are learning something from the course.  Adams and Wieman suggest that it is important for a test to contain items which satisfy these two criteria, so that faculty can see that the test measures student learning that they value.

\subsubsection{Wave-Particle Duality}

\begin{figure}[htbp]
  \includegraphics[width=\columnwidth]{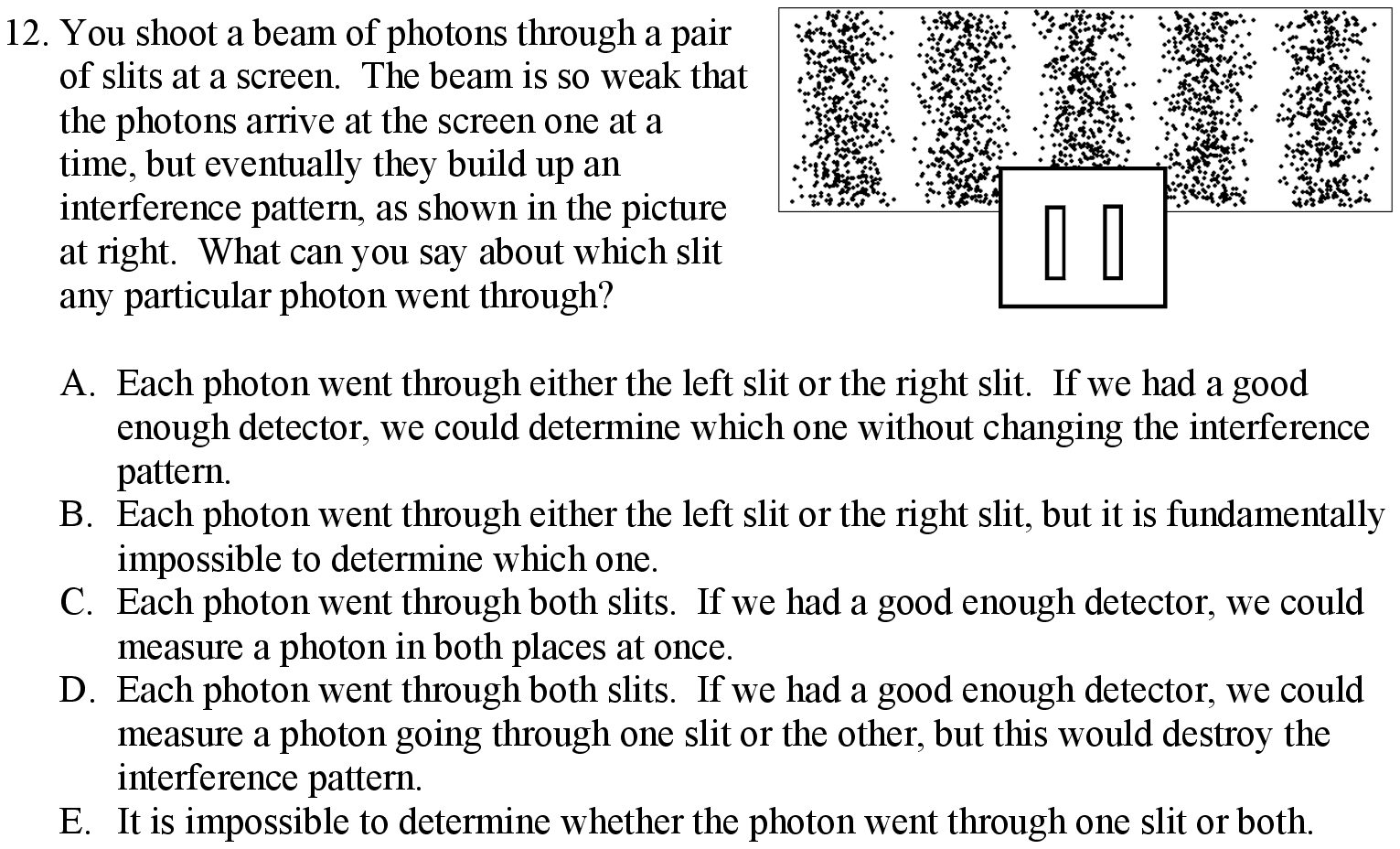}
  \caption{\label{Q12}Question 12 tests student understanding of the wave particle duality.}
\end{figure}

Wave-particle duality is another concept that many faculty in our interviews considered very important.  However, because this concept involves many subtle issues of interpretation, it is extremely difficult to test in a way in which experts can agree.~\cite{Dubson2009a}
Question 12, shown in Fig.~\ref{Q12}, is an attempt to address this important concept.  Although we have attempted many versions of this question, we still have not been able to find any version for which all physics professors agree with us on the ``correct'' answer.  The current version is an improvement over previous versions in that the majority of physics professors agree that our answer is the ``best'' answer of those given, even though many of them do not like it.

We believe that the correct answer is D, because the wave function must go through both slits to get an interference pattern, even though a detector can only measure a photon in one place.  If one solves the Schrodinger equation with a wave function going through only one slit, one does not get an interference pattern, so it is not correct to say that the photon went through only one slit, or even that we don't know whether it went through one slit or both.  In our wording, we have equated the photon with the wave function (or more precisely, with the electromagnetic wave, since a photon does not technically have a wave function), which we recognize is a particular choice of interpretation.  However, without this choice, there does not appear to be any wave-particle duality, and thus no way of testing this concept.

In interviews with 27 physics faculty, Dubson et al. found wide disagreement on the correct interpretation of the wave function, with $48\%$ of faculty interpreting the wave function as an information wave, $30\%$ interpreting it as a matter wave, and the remainder holding some kind of mixed view.~\cite{Dubson2009a}  Given these results, it is not surprising that we have had difficulty finding faculty consensus on any version of this question.  Our wording favors a matter wave interpretation, as defined by Dubson et al.  Faculty who prefer an information wave interpretation have argued that there is no correct answer, because while it is fair to say that the \emph{wave function} goes through both slits, one should not say anything at all about the \emph{photon}.  Unfortunately, changing the wording of the question to satisfy these faculty would eliminate the fundamental issue of wave-particle duality from the question, making it merely about the mathematical details of the wave function.

Another problem with interpretation with this question is that the small minority of faculty who prefer Bohm's interpretation of quantum mechanics choose B as the correct answer.  Further, a small minority of physics professors argue for answer E, either due to an agnosticism on interpretation or a lack of understanding of the implications of Bell's Inequality.

In interviews, on the other hand, modern physics students never argue for B or E due to deep issues of interpretation, but only due to a clear misunderstanding of the implications of quantum mechanics.  Further, although some previous versions of the question were difficult for students to interpret, the current version does not cause problems for students in terms of simply understanding what the question is asking.  Many students state answer D almost verbatim even before reading the options, so it accurately reflects the way that students think about this question.

\subsection{Questions designed to elicit student difficulties}

Questions designed to elicit specific student difficulties cited in the literature have been problematic from the perspective of faculty buy-in.  We have found that most faculty do not recognize the importance of such questions, either because they do not believe the underlying concept is important, or because they do not recognize the seriousness of the student misunderstanding.  Below we discuss several examples of such questions and the issues that they have raised.

\subsubsection{Belief that particles travel along sinusoidal paths}
Question 4, which asks whether photons travel along sinusoidal paths, began as a distracter for a question on the meaning of the wave-particle duality.  After seeing how often students in interviews responded that both photons and electrons do travel along sinusoidal paths, we discovered that many other researchers have reported on this difficulty~\cite{Steinberg1996a,Muller2002a,Olsen2002a,Knight2004b,Falk2004a}.  This particular student difficulty is unlike many of the difficulties reported in introductory physics in that it is not due to prior intuition, but due to instruction.  This is the only question where students sometimes do worse (often much worse) on the post-test than on the pre-test.  An example that illustrates how students might get this idea from instruction can be seen from an interview with a student who had just completed a modern physics course for physics majors.  He adamantly agreed with the statement that photons travel in sinusoidal paths, saying that his professor was always saying, ``photons move like this,'' and moving his index finger in a sinusoidal path.

This question has been problematic from the perspective of faculty buy-in.  Most faculty members who have reviewed the survey have not understood the purpose of this question because it does not occur to them that students might actually be thinking of photons and electrons as moving along sinusoidal paths.  Furthermore, faculty members, graduate students, and advanced undergraduates often pick the ``wrong'' answer for this question because they find the literal statement so nonsensical that they automatically translate it in their heads to something like, ``the wave function of the photon is sinusoidal,'' which is a true statement.  (We added the qualifying phrase ``in the absence of external forces'' because several graduate students pointed out that electrons could travel in sinusoidal paths if the right forces were applied, for example in an undulator.)

\begin{figure}[htbp]
  \includegraphics[width=\columnwidth]{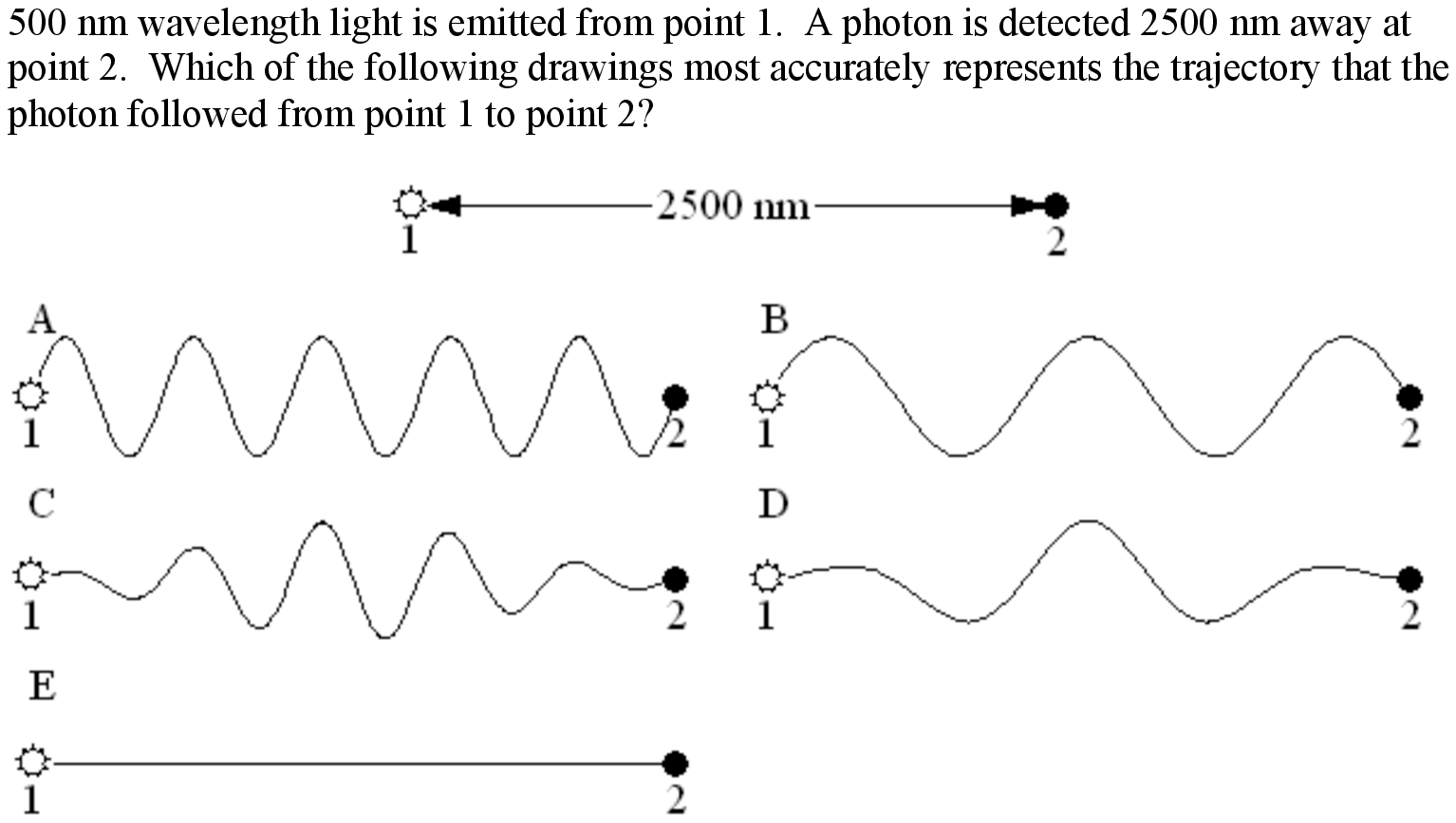}
  \caption{\label{sinusoidal}An alternative version of question 4, taken from Knight.  This version does not work well because too many students do not notice the word ``trajectory.''}
\end{figure}

However, the modern physics students we have interviewed almost never misinterpret the question.  Of the 46 modern students we interviewed on this question, only 2 misinterpreted the statement to mean that the wave function or electromagnetic field is sinusoidal and therefore gave the incorrect answer even though they understood the concept.  The remainder of the students, regardless of whether they thought the statement was true or false, interpreted it as meaning that electrons literally travel in sinusoidal paths.  Many students moved their fingers or hands in a sinusoidal motion to illustrate the path.  Thus, while the question is not valid for advanced students and experts due to interpretation problems, it does appear to be valid for modern physics students.  We have attempted many different wordings of the question to in order to avoid interpretation problems, but have not found any that work better than the current wording.  In some versions of the QMCS, we tested out the question shown in Figure~\ref{sinusoidal}, taken from Knight~\cite{Knight2004b}, which gets at the same concept with pictures rather than words.  We found that this question was not valid even for modern physics students, as many students who correctly answered question 4 answered this question incorrectly, reporting that they missed the word ``trajectory'' or that they saw numbers and automatically went into calculation mode without reading the rest of the question.

\subsubsection{Difficulty understanding relationship of amplitude and wavelength to potential}

\begin{figure}[htbp]
  \includegraphics[width=\columnwidth]{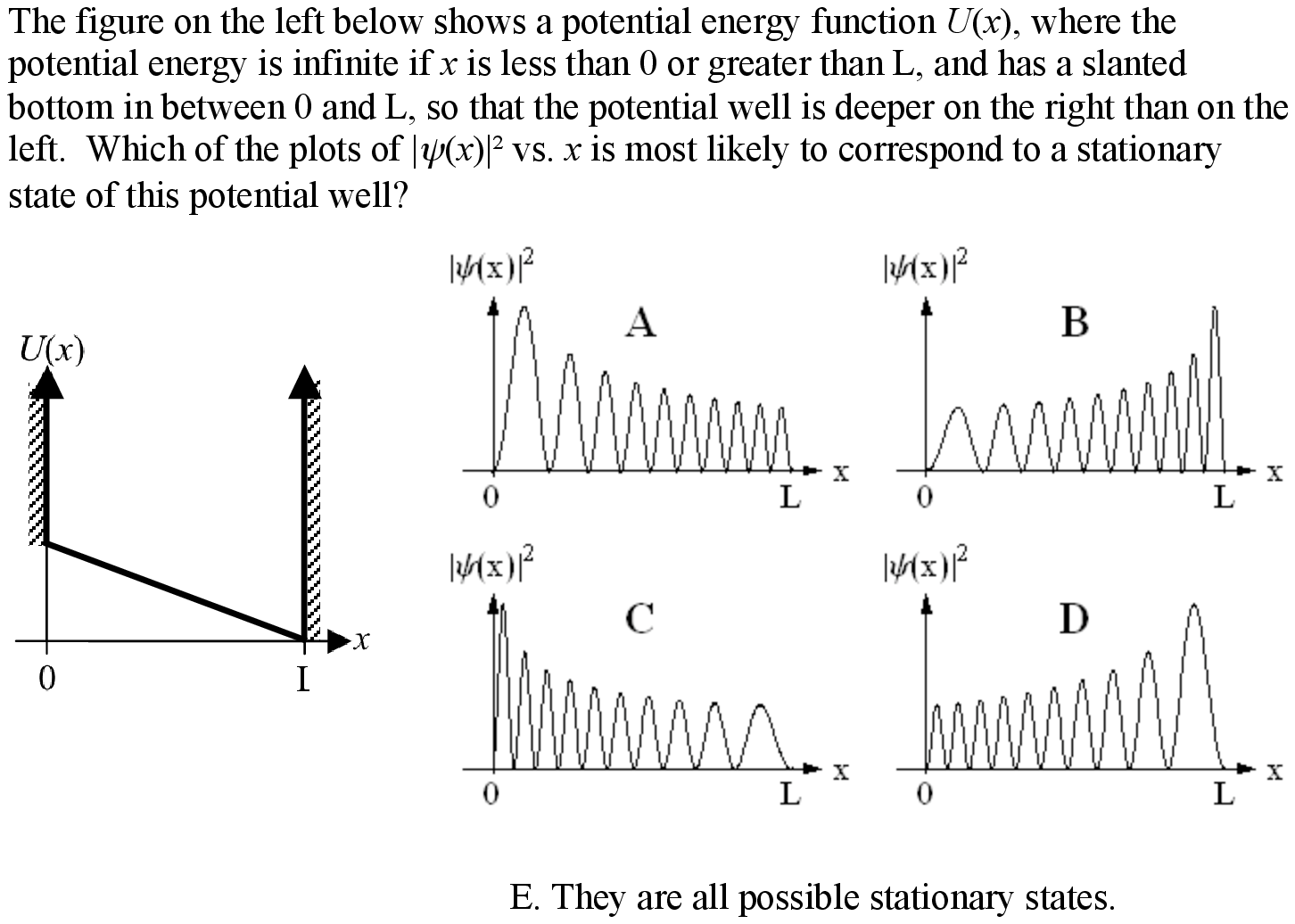}
  \caption{\label{slanted}The original version of question 10, taken from the QMVI.  In the current version this question has been adapted in order to make it easier, by including only answers with larger magnitude on the left side.  The correct answer is A.}
\end{figure}

There is an extensive literature of physics education research demonstrating that students have difficulty understanding the qualitative relationships between the potential energy function and both the local wavelength and the amplitude of the wave function.~\cite{Ambrose1999a,Bao1999a,Sadaghiani2005a}  However, after reviewing textbooks and syllabi, we suspected that this difficulty might be due simply to a lack of coverage in standard courses, rather than to any deep underlying conceptual difficulty.  With notable exceptions in the textbooks of French and Taylor~\cite{French1978a} and of Robinett~\cite{Robinett1997a}, these concepts do not appear to be covered in any depth in standard textbooks or syllabi of either modern physics or quantum mechanics courses.

The original version of question 10, taken from the QMVI, shown in Figure~\ref{slanted}, was included in early versions of the QMCS in order to test these difficulties.  The original question required understanding two separate concepts: that the local deBroglie wavelength is shorter in regions of lower potential energy (because KE = E - PE is larger), and that the amplitude is larger in regions of lower potential energy (because larger KE means the particle spends less time there).  Each concept can be evaluated separately by looking at the percentage of students who answered either A or B (wavelength correct) and the percentage of students who answered either A or C (amplitude correct).

\begin{figure}[htbp]
  \includegraphics[width=\columnwidth]{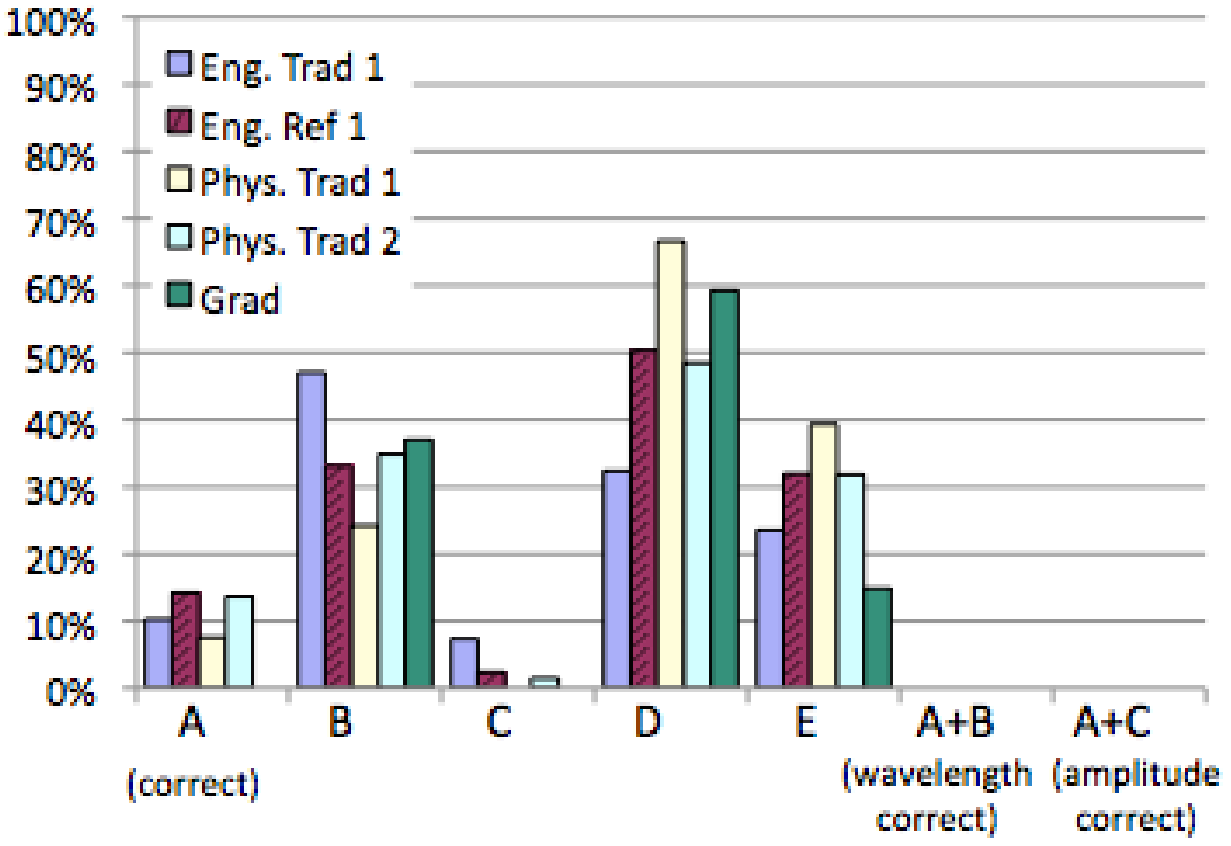}
  \caption{\label{slantedresponses}Responses to the original slanted well question shown in Fig.~\ref{slanted} for students in four modern physics classes and for graduate students.}
\end{figure}

Our results suggest that these concepts are not well-understood by students or by experts.  Fig.~\ref{slantedresponses} shows that modern physics students perform extremely poorly on this question, as one might expect from the results of previous research.  An unexpected result was the poor performance of experts.  Only 15\% of graduate students were able to answer this question correctly, on par with modern physics students, and many faculty, including some very well-known and respected physicists, have answered it incorrectly or said that they do not know how to answer it.

In spite of this poor performance, interview results suggest that the underlying concepts are not particularly difficult.  When the interviewer reviewed the answers with students at the end of interviews, students who had initially gotten it wrong were almost always able to figure out the correct answer after a few leading questions from the interviewer (e.g. Is the kinetic energy larger where the potential energy is low or where it is high?  If you have a pendulum swinging back and forth, does it spend more time in the regions where it moves quickly or in the regions where it moves slowly?).  For most other questions on the QMCS, if students failed to answer them correctly on their own, it required much more extensive guidance to help them understand the correct answer.

Fig.~\ref{slantedresponses} shows that students in one modern physics course scored significantly higher on this question than both other modern physics students and graduate students.  In interviews with nine students from this course, we learned that the professor had emphasized the concepts in this question in lecture.  Eight out of nine students mentioned specific details from his lectures, such as that he frequently said that ``the curvature encodes the kinetic energy" and that he used a pendulum example to explain why the amplitude is higher when the potential energy is higher.  Even students who answered the question incorrectly remembered these lectures, but misremembered or misunderstood some aspect of them.  Further, the students who answered correctly were transferring knowledge to a new situation: One student said in an interview that the professor had shown an example with a step in the well, not a slanted well, and that he was not sure that what he had said applied to a continuously varying potential but he thought it probably did.  Students from other courses did not mention lectures when answering this question, although one of these students got the wavelength correct because he clearly remembered a homework problem designed to address this concept, which had been the only such instruction in the entire course.

Thus, our results indicate that students' inability to answer this question is due mainly to a lack of exposure, and it may be possible to remedy this with minimal instruction.

The amplitude concept in this question appears to be more difficult than the wavelength concept.  This can be seen from the responses shown in Fig.~\ref{slantedresponses}, and from our student interviews and discussions with faculty, in which both groups had trouble with both concepts, but much more difficulty with the amplitude.  While grad students performed as poorly as modern physics students on the question as a whole, they were more likely than most groups of modern physics students to give answers that included the correct wavelength.  In our review of textbooks, we found that most modern physics textbook have at least a homework question or two that address the qualitative relationship between wavelength and potential, but do not address the qualitative relationship between amplitude and potential at all.  

Because the concept of amplitude addressed in this question is so poorly understood even by experts and is not covered in most textbooks, we modified the question to the current question 10, which tests only student understanding of wavelength, and not amplitude.  In interviews with the current version, students are often bothered by the fact that the amplitude is larger on the left, but they answer the question based on their understanding of the wavelength.

The modified question is still the most difficult question on the QMCS, shows little discrimination between weak and strong students, and shows no improvement between pretest and posttest.  In spite of these issues, we have retained question 10 because it successfully illustrates that the relationship between wavelength and potential is not well understood by most students.  Furthermore, as demonstrated by the one course in which we saw improvement on the original version of this question, it appears that it is possible to make significant gains in understanding of this concept with minimal but targeted instruction, and that this question can be used to test such instruction.  We encourage others to use this question to test the effectiveness of different types of instruction for teaching this concept.

\subsubsection{Belief that reflection and transmission are due to a range of energies}

\begin{figure}[htbp]
  \includegraphics[width=\columnwidth]{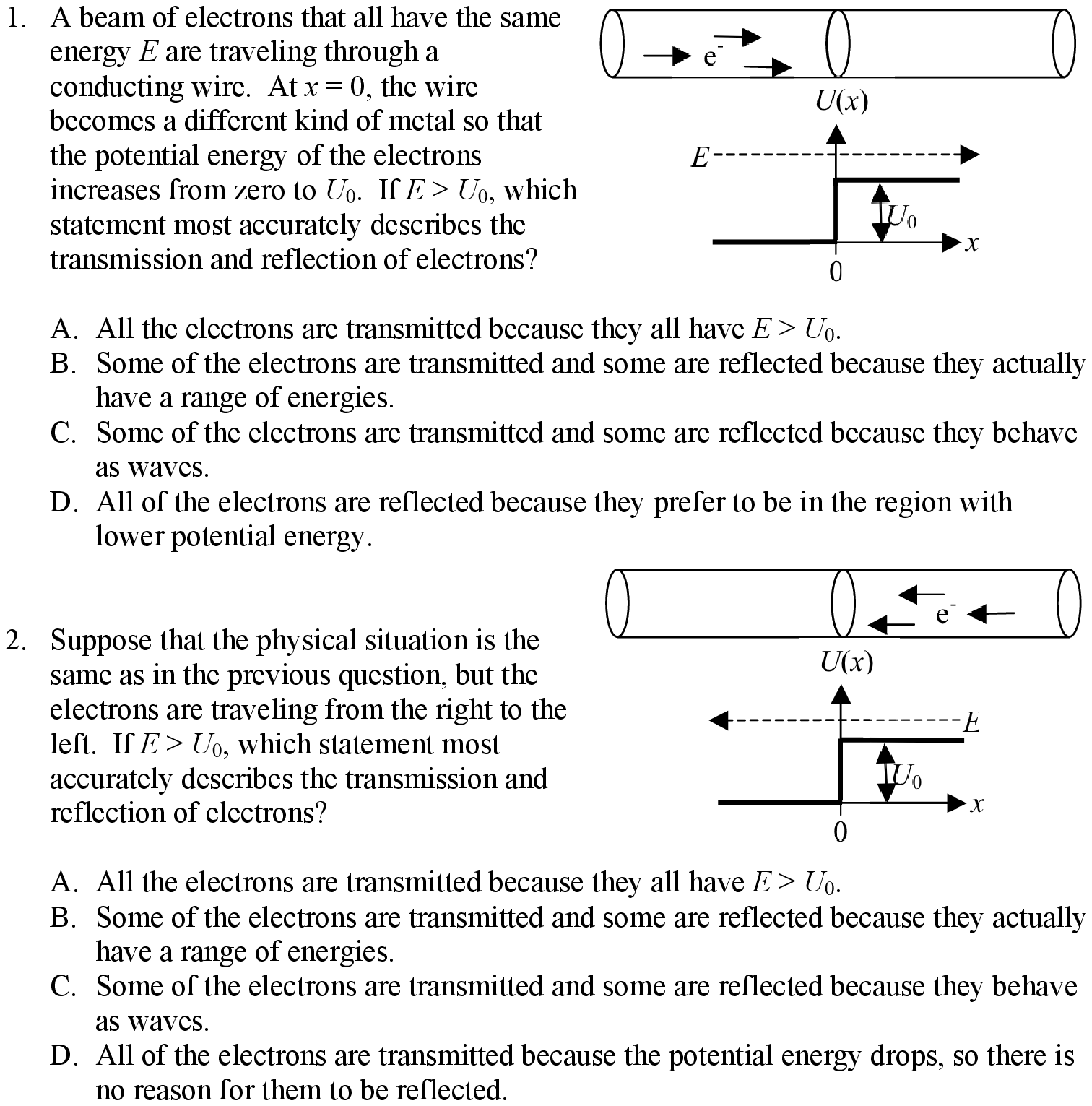}
  \caption{\label{RT}Two questions designed to elicit the belief that reflection and transmission are due to a range of energies, adapted from Ref.~\onlinecite{Ambrose1999a}.  The correct answer to both questions is C.}
\end{figure}

Figure~\ref{RT} shows two questions designed to elicit the belief that reflection and transmission are caused by particles having a range of energies rather due to the wave nature of matter, another student difficulty previously discussed in the physics education research literature.~\cite{Ambrose1999a,Bao1999a,Domert2005a}  The results of interviews with the first question have been discussed elsewhere~\cite{McKagan2006a}, and will be summarized only briefly here: Students often believe that reflection and transmission are caused by a range of energies.  Students with this belief may select answer B, which characterizes this belief, or A, stating that in this case there will not be reflection because there is not a range of energies.  This is consistent with results that have been reported by other researchers.

However, while other researchers have attributed this error to ``treating electrons as particles rather than waves,'' our results are not consistent with this interpretation.  Rather, we have consistently seen students make this error while clearly describing electrons as waves.  As a result of this observation, we started asking students who had answered this question incorrectly in interviews to answer additional questions about what would happen to a wave pulse traveling along a rope that is heavier on one side than the other, and/or a beam of light traveling from air to glass and vice versa.  We found that many students answered these questions incorrectly as well, indicating that the problem is not specific to quantum mechanics.  Many students do not appear to view the behavior of transmitting only when there is enough energy as a property of particles, but of common sense.  Thus, the problem appears to be that students do not understand wave behavior, rather than that they view electrons as particles. This distinction has important pedagogical consequences because it means that simply emphasizing the wave nature of electrons is not sufficient to address this error.

A further problem with these questions is that we have found that the ``correct'' answer (C) is somewhat unsatisfying to both faculty and students who do \emph{not} have the incorrect belief that partial reflection is caused by a range of energies.  Students sometimes reported that they chose C because all the other answers were clearly wrong, but it didn't really seem like an explanation.  For example, one student gave the following explanation for his answer: ``I was pretty sure that some of the electrons were reflected and some were transmitted, and... somehow that has to do with them behaving as waves.  And since that was all the question asked, I was like well, that's about the level I know.''  Another student said, ``Because they behave as waves?  That doesn't really do a good job of explaining anything!''  Several faculty members also reported discomfort with the answer, arguing that while it was true, it wasn't really an explanation.  However, neither faculty nor students were able to offer a more satisfactory explanation.  The original wording, in an early version that was tested with only 7 students, was ``because of the interaction with the potential step'' rather than ``because they behave as waves.''  This wording was even more confusing to students, who were unable to interpret the meaning of the statement at all.

We have removed these questions from Version 2.0 for two reasons.  First, we found in discussions with faculty that a substantial fraction of them did not cover the concepts addressed in these questions in their modern physics classes.  Because this topic is not universally covered, the questions are not useful for comparing different courses because they measure whether the topic was covered rather than the quality of instruction.  Second, our goal in the QMCS is to measure \emph{conceptual} understanding of quantum mechanics, and we found in interviews that answering these questions does not require conceptual understanding.  Students who answered correctly usually did so from memory, and did not display significantly deeper conceptual understanding than those who did not.

\subsubsection{Difficulty drawing correct wave functions for tunneling planes waves}
Another difficulty commonly cited in the literature is the inability to draw correct wave functions for tunneling plane waves.~\cite{Ambrose1999a,Bao1999a,Morgan2004a,Wittmann2005a}  Commonly cited mistakes are drawing the transmitted part of the wave function with an offset or a smaller frequency, and drawing the tunneling part of the wave function as missing or sinusoidal.  As discussed in detail in Ref.~\onlinecite{McKagan2008c}, early versions of the QMCS included a question designed to address this difficulty, but this question was eliminated because we saw in interviews that answering it correctly did not require any conceptual understanding, only memory.

\subsubsection{Belief that energy is lost in tunneling}
Students often believe that energy is lost in tunneling.~\cite{Ambrose1999a,Bao1999a,Morgan2004a,Wittmann2005a,Falk2004a,Domert2005a,McKagan2006a}  This difficulty, like the previous two difficulties, is quite difficult to assess in a conceptual way.  The original question designed to address this difficulty, discussed in great detail in Refs. \onlinecite{McKagan2006a} and \onlinecite{McKagan2008c}, does require conceptual understanding to answer, but we found that it is measuring too many things at once and, especially for students in reformed courses, answering it incorrectly does not necessarily indicate a belief that energy is lost in tunneling.  To ensure that the question is measuring what we intend, we have replaced it with question 7, which asks more directly about energy loss in tunneling.  The new question requires much less conceptual understanding, but it is easier to interpret the results.  We have retained it because this is an important concept that students often have trouble understanding.

\section{Suggestions for using the QMCS}

The QMCS is intended to be used as a formative assessment tool for faculty to measure the effectiveness of different teaching methods at improving students' conceptual understanding of quantum mechanics, and to use such measurements to improve their teaching.  We believe that it is useful for this purpose in modern physics courses, and it may be useful for this purpose in junior quantum courses.

Version 2.0 of the QMCS should take about 20 minutes for students to complete.

We recommend giving the QMCS in class as a paper test with scantron answer sheets.  To protect the security of the test, it is important to collect all tests and answer sheets, and not to post results, so that future students cannot use the test to study.

We have also administered the QMCS online on a password-protected site that is available for a limited time with security measures to prevent copying and pasting.  We have found that this method works well for graduate students, who are typically motivated by the material and find the test intrinsically interesting.  For modern physics students, administering the test online gives unacceptably low response rates.  Online administration might work for students in a junior quantum course, but we have not tried it.

To reduce students' motivation to keep the test to study, we do not recommend basing any part of their course grades on their scores on the QMCS.  We do recommend offering a small amount of participation credit to encourage them to take it.

\subsection{Modern Physics Courses}
The QMCS has been most thoroughly tested at this level, with extensive interviews and statistical analysis of results with students in a wide variety of courses at the University of Colorado, including courses for engineering majors and physics majors, and traditional and reformed courses.

We recommend giving the QMCS as a posttest only in modern physics classes.  Students cannot be expected to know any of the content before such a class, and we have found that giving the QMCS as a pretest to modern physics students is extremely demoralizing.

We have found that a good way to motivate students to take the QMCS seriously is to give it on the second-to-last day of class as a ``final exam prep".  We then spend the last day of class going over the answers to questions on which students did particularly poorly.  Students report that they find it useful when administered in this way.

We encourage others to administer the QMCS in modern physics courses and to use the results to inform their teaching, as well as to publish them to inform the broader community.

\subsection{Junior Quantum Courses}
Very little testing of the QMCS has been done at the junior quantum level.  Our preliminary results suggest that it is a useful test at this level, and should be given as a pretest and a posttest in order to examine learning gains.  However, further research is needed to determine typical scores that should be expected and to test the validity of the questions for students at this level.  We encourage others to administer the QMCS in junior quantum courses, to use the results to inform their teaching, and to report their results.

\subsection{Graduate Quantum Courses}
The QMCS has been administered to dozens of graduate students in an online format allowing for optional comments on each question.  We have used these comments to refine the test.

The QMCS does distinguish between students in more or less competitive graduate programs, and it may be useful for assessing graduate students' understanding of the most basic conceptual ideas in quantum mechanics.  However, when we have administered it as a pretest and posttest in graduate courses, we find zero gains, even for students with very low pretest scores.  The QMCS does not test any of the more advanced ideas that students are expected to learn in graduate quantum mechanics, and thus is not useful for assessing such courses.

\section{Suggestions for further research}
We have studied the validity and application of the QMCS extensively in modern physics and graduate quantum courses.  However, because of the particular difficulties associated with conceptual assessment of quantum mechanics discussed in Section IIIE, further research is needed to develop ideal assessment tools for this subject.  We encourage others to test the QMCS at different institutions and levels, as well as to develop new questions.~\cite{Deslauriers2010a}

\section{Acknowledgments}
We thank the Physics Education Research Group at the University of Colorado for extensive feedback on all aspects of the development of this survey.  This work was supported by the NSF and the University of Colorado Science Education Initiative.

\appendix

\section{Learning Goals for QMCS}
Below are the specific learning goals for each question on Version 2.0 of the QMCS, with the general concepts tested by each question in parentheses:
\begin{enumerate}
  \setlength{\itemsep}{1pt}
  \setlength{\parskip}{0pt}
  \setlength{\parsep}{0pt}
\item Apply the concept of quantized energy levels to determine the relationship between energy and wavelength in an atomic transition (wave-particle duality, quantization of states)
\item Interpret the solutions of the Schrodinger Equation to recognize that electron energy levels are spread out in space (wave-particle duality, wave function and probability, uncertainty principle, operators and observables, measurement)
\item Apply the concept of quantized energy levels to calculate the energies of photons emitted in transitions (quantization of states)
\item Recognize that the wave-particle duality does NOT imply that particles move in wave-like trajectories. (wave function and probability, wave-particle duality)
\item Apply the deBroglie relationship to determine the wavelength of a particle from its properties of motion (wave function and probability, wave-particle duality)
\item Apply the uncertainty principle to determine the possible outcomes of a measurement (uncertainty principle, superposition, operators and observables, measurement)
\item Recognize that energy is not lost in tunneling (tunneling)
\item Use the Schrodinger Equation to determine when energy is quantized (Schrodinger Equation, quantization of states)
\item Use the Schrodinger Equation to determine when energy is not quantized (Schrodinger Equation, quantization of states)
\item Determine qualitatively the shapes of solutions to the Schrodinger Equation (Schrodinger Equation)
\item Given a wave function, determine the probability distribution (wave function and probability)
\item Describe the model that quantum mechanics gives to explain the double slit experiment (wave function and probability, measurement)
\end{enumerate}

\bibliography{../bibliographies/PER2}
\bibliographystyle{apsrev4-1}

\end{document}